\documentclass{iaus}
\usepackage{graphicx}

\title[Abundance gradients in the galactic disk] 
{Abundance gradients in the galactic disk: \\ space and time variations}

\author[Walter J. Maciel \& Roberto D. D. Costa]   
{Walter J. Maciel \and Roberto D. D. Costa}

\affiliation{Astronomy Department, University of S\~ao Paulo,\\ Rua do Mat\~ao 1226,
05508-900, S\~ao Paulo SP, Brazil \\ email: {\tt maciel@astro.iag.usp.br} 
\\email: {\tt roberto@astro.iag.usp.br}}

\pubyear{2008}
\volume{254}  
\pagerange{1--6}
\jname{The Galaxy Disk in Cosmological Context}
\editors{J. Andersen, J. Bland-Hawthorn \& B. Nordstr\"om, eds.}
\begin{document}

\maketitle

\begin{abstract}
Recent work on abundance gradients have focussed not only on their magnitudes, but 
also on their spatial and temporal variations. In this work, we analyze the behaviour 
of radial abundance gradients in the galactic disk giving special emphasis on these 
variations. The data used includes planetary nebulae and objects in different age brackets, 
namely open clusters, HII regions, cepheid variables and stars in OB associations. We find 
evidences for a space variation of the radial gradients as measured  for element ratios such 
as O/H, S/H, Ne/H, Ar/H and [Fe/H], in the sense that the gradients tend to flatten out at 
large galactocentric distances. Moreover,  near the bulge-disk interface a steep decrease 
in the abundances is observed. The time evolution of the gradients is also evaluated on the 
basis of approximate ages attributed to the central stars of planetary nebulae and open 
cluster stars. It is concluded that the available data is consistent with a time flattening 
of the gradients during the last 6 to 8 Gyr, a time interval in which the age determinations 
are probably more accurate.
\end{abstract}

\firstsection 
\section{Introduction}

Radial abundance gradients in the galactic disk are among the main constraints 
of chemical evolution models. They can be derived from a variety of objects, 
such as HII regions, planetary nebulae (PN), open clusters and young stars. 
Recent work on abundance gradients have focussed on (i) their magnitudes, 
(ii) their spatial variations and (iii) their temporal evolution, so that the 
number of derived observational constraints has been considerably increased. In 
this work, we analyze the behaviour of radial abundance gradients using a large 
sample of PN for which the ages of the central stars (CSPN) have been estimated. 
Such data is complemented with other objects in different age brackets, 
such as open clusters, HII regions, cepheid variables and stars in OB associations. 
We focus on the possible space variations of the gradients as measured  for elements 
such as O, S, Ne, Ar and Fe at large galactocentric distances and near the 
bulge-disk interface, as well as on the time evolution of the gradients based on 
approximate ages attributed to the CSPN and open cluster stars. 

\section{Space variations of the gradients}

Space variations of the radial abundance gradients have been suggested in the literature, 
comprising basically three different regions: (i) a change of slope at large galactocentric 
distances, $R > 10\,$ kpc, assuming the solar distance to be in the range $R_0 = 7-8\,$ kpc; 
(ii) a flattening in the inner disk, roughly corresponding to $R < 4\,$ kpc; and (iii) a 
flattening or a discontinuity  near the solar circle. 

\begin{figure}
   \centering
   \includegraphics[angle=-90,width=8cm]{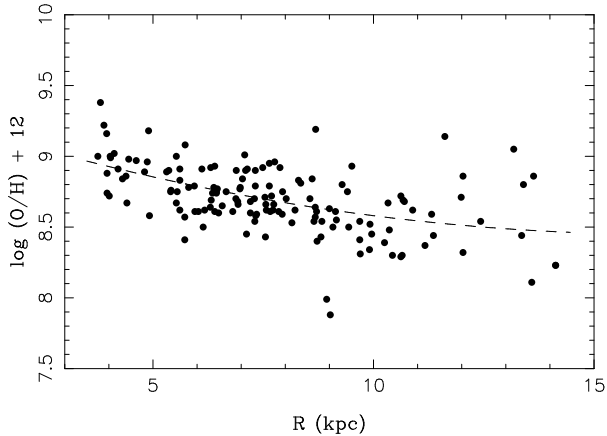}
      \caption{The O/H radial abundance gradient in the galactic disk.
              }
         \label{fig1}
   \end{figure}

(i) Nebulae in the anticentre direction were the subject of a project developed by the 
IAG/USP group in order to establish the behaviour of the radial gradient at large 
galactocentric distances (\cite[Costa et al. (2004)]{costa2004}). A large sample was 
observed with the 1.5m ESO telescope, and the results suggest that the gradient flattens 
out beyond $R > 10\,$ kpc, as shown by Fig.\,\ref{fig1}. The dashed line shows a 
second order gradient to stress the flattening at high $R$. These results were partially 
anticipated on the basis of a smaller sample by \cite[Maciel and Quireza (1999)] {mq99}, 
and have been confirmed by independent work on HII regions by \cite[V\'\i lchez \& Esteban 
(1996)]{vilchez}. However, an approximately constant gradient was found by some recent work 
on HII regions by \cite[Deharveng et al. (2000)]{deharveng}, \cite[Quireza et al. 
(2006)]{quireza}, \cite[Rudolph et al. (2006)]{rudolph} and \cite[Rood et al. (2007)]{rood}. 
The flattening observed in the anticentre direction from PN data can also be noticed in the 
open cluster survey by the Bologna Open Cluster Chemical Evolution Project (BOCCE, see for 
example \cite[Bragaglia and Tosi (2006)]{bragaglia}, \cite[Sestito et al. (2006)]{sestito2006}, 
\cite[Sestito et al. (2007)]{sestito2007}), based on high resolution spectroscopy, suggesting 
that some flattening in the [Fe/H] gradient can be observed for $R \sim 12\,$ kpc. This 
conclusion is also supported by a very recent work based on a sample of galactic cepheids 
(\cite[Lemasle et al. (2008)]{lamasle}), in which a higher dispersion was found in the 
outer disk, as in Fig.\,\ref{fig1}. Other observational evidences of the flattening of the 
gradients at large galactocentric distances are discussed by \cite[Cescutti et al. (2007)]{cescutti}, 
based on open clusters, cepheids and red giants, for the O/H and [Fe/H] ratios, and other elements.

(ii) In a recent work, about 50 PN in the galactic bulge have been analyzed by \cite[Cavichia 
(2008)]{cavichia}, (see also Costa et al., this conference), in an investigation of the 
interface between the galactic disk and bulge. Apart from obtaining accurate abundances for 
these nebulae, an effort was made to separate the two populations based on their chemical 
composition. It was found that the abundance trend presented by disk objects is {\it not} 
maintained as the bulge region is approached, and the bulge population has lower average 
abundances than predicted by a simple extrapolation of the disk gradient. In this case,
Group I, or disk objects, have systematically larger abundances than Group II, or bulge
nebulae. The galactocentric distance that represents the interface of the two populations, 
obtained through a Kolmogorov-Smirnov test, is in the range $R \sim 1.5-3.0\,$ kpc.  

Recently, \cite[Pottasch \& Bernard-Salas (2006)]{pottasch} presented a summary of PN 
abundances homogeneously determined from ISO measurements, and mid-infrared results of bulge 
PN from Spitzer have also been recently discussed by Gutenkunst et al. (2008). These results 
can be used  to analyze the behaviour of PN abundances near the galactic bulge, as compared 
with the inner disk objects, thus contributing to the study of the space variations of the 
abundance gradients. On the basis of bulge nebulae with Spitzer data, it can be seen from 
the data by \cite[Gutenkunst et al. (2008)]{gutenkunst} that the bulge PN abundances of O, 
Ar, Ne, and S do not follow the abundance gradient observed in the disk, in the sense that 
these abundances are systematically lower than predicted by the disk gradient. These results 
agree very well with our own results obtained for PN in the bulge-disk interface, as 
presented by Costa et al. (this conference) and with previous suggestions in the literature,
such as \cite[G\'orny et al. (2004)]{gorny}.

(iii) There are some observational evidences in favour of space variations of the gradients 
near the solar circle, in a dividing line located about $R \simeq 10\,$ kpc, especially based 
on cepheid variables (\cite[Andrievsky et al. (2004)]{andrievsky}, however, see \cite[Lemasle 
et al. (2008)]{lemasle}) and open clusters (\cite[Twarog et al. (1997)]{twarog}, \cite[Corder 
\& Twarog (2001)]{corder}, \cite[Yong et al. (2005)]{yong}). PN cannot at present contribute 
decisively to this subject, as the uncertainties in the abundances, distances and ages of 
their central stars prevent the knowledge of a detailed behaviour of the gradients in that 
region. 

\section{Time variations of the gradients}

\begin{figure}
   \centering
   \includegraphics[angle=-90,width=8cm]{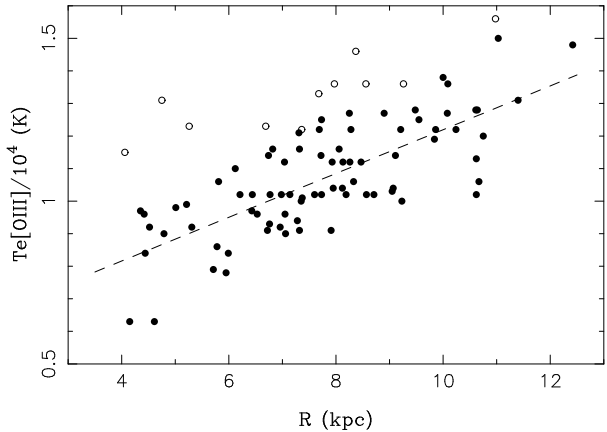}
      \caption{The PN electron temperature gradient from \cite[Maciel et al. 
       (2007)]{mkc2007}.
              }
         \label{fig2}
   \end{figure}

Recently, \cite[Quireza et al. (2006)]{quireza} presented a detailed investigation of over a 
hundred HII regions within about 17 kpc of the galactic centre with accurate determinations of 
the electron temperatures from radio recombination lines. From this study, an average gradient 
of $dT_e/dR \simeq 290\,$K/kpc was obtained.  On the other hand, \cite[Maciel et al. 
(2007)]{mqc2007} investigated the electron temperature gradient of PN in the galactic disk as 
measured by [OIII] electron temperatures, and compared their results with the gradient from 
HII regions derived from radio recombination lines. Since the abundance of the main coolants 
is inversely correlated with the temperature, the temperature gradient can be considered as a 
mirror image of the abundance gradient, so that an analysis of the former can be used in order 
to confirm any time variations in the latter. A sample of PN central stars with ages 
averaging 4--5 Gyr was selected, and their nebular [OIII] electron temperatures were plotted 
against the galactocentric distances, as shown in Fig.\,\ref{fig2}. The empty circles 
represent CSPN with extremely high temperatures, which show larger deviations from the trend.
The average gradient (dashed line) is about $dT_e/dR \simeq 670\,$K/kpc, depending slightly 
on the adopted distance scale. This result not only confirms a strong abundance gradient, 
but also gives an important and independent confirmation of the time variation of the 
gradients, as the HII regions are much younger than the PN sample. Since the observed PN 
gradient is steeper than that of the HII regions, we have an independent evidence that the 
gradients have flattened out during the last 5 Gyr, approximately. 

\begin{figure}
   \centering
   \includegraphics[angle=-90,width=8cm]{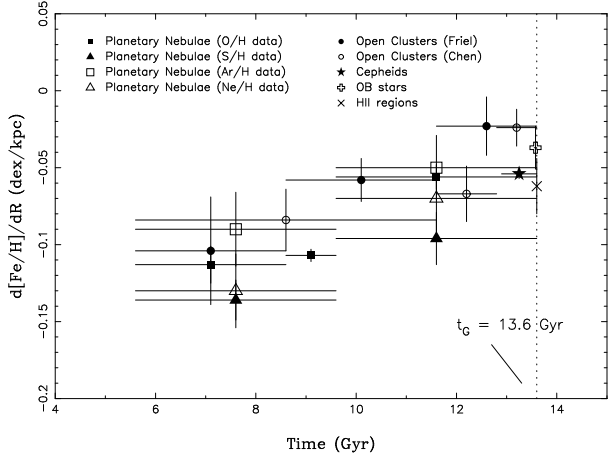}
      \caption{Time variation of the converted [Fe/H] radial gradient from PN
      and other objects.
              }
         \label{fig3}
   \end{figure}

In a series of papers, \cite[Maciel et al. (2003)]{mcu2003}, \cite[Maciel et al. 
(2005)]{mlc2005}, and \cite[Maciel et al. (2006)]{mlc2006} studied the time variation of the 
abundance gradients in the galactic disk by adopting a new approach regarding CSPN. While most 
investigations on this topic based on PN assumed the traditional Peimbert types as defining 
different classes of PN, Maciel et al. tried to obtain individual ages of all stars, thus 
avoiding the main disadvantage of the Peimbert system, namely, the fact that objects with 
different masses, or different ages, may be assigned to the same type. As a consequence, it 
was possible to calculate different gradients according to the age of the central stars, so 
that the time evolution of the radial gradients could be determined. In practice, several 
difficulties arise, basically due to the fact that the sample sizes are limited and the ages 
are still uncertain. Nevertheless, reliable {\it relative} gradients can be estimated, as 
shown in Fig.\,\ref{fig3}, where the abundance gradients, converted to [Fe/H] gradients (cf. 
\cite[Maciel (2002)]{maciel2002}), are plotted against time, adopting a galactic age of 13.6 
Gyr.  These results are consistent with some flattening of the gradients during the time 
interval when the data is more reliable, namely the last 6 to 8 Gyr. 

An analysis of recent determinations of abundance gradients from PN in the literature shows 
that there is no general consensus regarding their average values, as well as the space and 
time variations. While \cite[Pottasch \&  Bernard-Salas (2006)]{pottasch} and  
\cite[Gutenkunst et al. (2008)]{gutenkunst} obtain stronger gradients similar to those 
derived by our group, \cite[Perinotto \& Morbidelli (2006)]{perinotto} find evidences for 
weaker gradients. In fact, some investigators failed to determine measurable abundance 
gradients from PN, which can be due to several reasons: (i) The gradients are generally small, 
so that the total variation in the abundances depend on the considered range in the 
galactocentric distances; (ii) The associated uncertainties of the abundances can be large, 
which makes the detection of the gradients more difficult; (iii) The distances of PN are 
poorly known; (iv) Assuming that the gradients change with time, placing objects with different 
ages on the same plot would obviously increase the dispersion, leading to essentially flat 
gradients. We made some simulations adopting three groups of objecs with 
progressively older ages and steeper gradients. If all objects are put together, the main 
result is a blurring of the gradients, leading to a flatter gradient, the magnitude of which 
would depend on the proportion of stars in each group.

\cite[Perinotto \& Morbidelli (2006)]{perinotto} derived O/H abundance gradients from disk PN 
using a recent and homogeneous set of chemical abundances based on different distance scales. 
The derived gradients are generally small, under $-$0.04 dex/kpc, but no effort was made in 
order to consider the objects according to group ages.  The same approach was taken years ago 
by \cite[Maciel \& K\"oppen (1994)]{mk94} and \cite[Maciel \& Quireza (1999)]{mq99} and 
abandoned in view of its shortcomings. However, even the detailed analysis by \cite[Perinotto 
\& Morbidelli (2006)]{perinotto} shows clearly that some time variation of the gradients is 
to be expected. Their so-called Set A, considered as a reliable homogeneous set of abundances, 
has a gradient of $-$0.017 dex/kpc, which is clearly steeper than the data for Set B. This is 
a control sample, or a mixed sample in which any existing gradient would probably have been 
erased. A similar result (that is, no gradient) is also observed in the subsample of Type I PN, 
which have massive central stars, and therefore are relatively younger objects compared to the 
remaining stars in Set A. In both cases, that is, Set B and the Type I PN of Set A,  
no gradients are found. For the Type II PN of Set A, which come in principle from less massive 
central stars, and therefore are expected to be older than Type I objects, the gradient found 
is steeper, about $-$0.025 dex/kpc, which suggests some time evolution of the gradients in the 
sense that older objects have steeper gradients. This is reinforced by the fact that Type III 
PN, which are expected to be older, show an even steeper gradient. Although all gradients 
obtained by \cite[Perinotto \& Morbidelli (2006)]{perinotto} are relatively flat compared 
with our disk sample, it is seen that their results also suggest some time flattening of the 
gradient, in agreement with our own results. 

In Fig.\,\ref{fig3} we show our PN data along with data from different types of objects. 
Regarding young objects, we include HII regions and OB stars, which are located essentially 
near the 13.6 Gyr line, as these are the youngest objects of all. A class of stars with 
slightly higher ages are the cepheid variables, for which extremely accurate abundances and 
distances have been determined. Since the average ages of these stars are lower than about 
700 Myr (see \cite[Maciel et al. (2005)]{mlc2005} for a detailed discussion), they contribute 
strongly to the accuracy of the present day gradient shown in Fig.\,\ref{fig3}. For the older 
objects, only the open clusters have an age distribution comparable to the PN central stars, 
so that these objects are especially important in the study of the time evolution of the 
gradients. As mentioned, there is presently no large sample of open clusters with 
homogeneously derived abundances and distances, but available data as given by \cite[Friel 
et al. (2002)]{friel} and especially the compilation by \cite[Chen et al. (2003)]{chen} 
strongly support the PN data, as can be seeen in Fig.\,\ref{fig3}. The recent high resolution 
data from the BOCCE project still does not have enough results to tackle this problem directly, 
but we can use the available data to make some estimates. From \cite[Bragaglia \& Tosi 
(2006)]{bragaglia}, \cite[Sestito et al. (2006)]{sestito2006}, and \cite[Sestito et al. 
(2007)]{sestito2007} we see that  at galactocentric distances smaller than about $R 
\sim 12\,$kpc, a gradient similar to the one derived by \cite[Friel et al. (2002)]{friel} 
is obtained, which is about $d$[Fe/H]$/dR \sim -0.06\,$dex/kpc. Adopting the disk Fe-O 
calibration by \cite[Maciel (2000)]{maciel2000}, we have approximately $d$(O/H)$/dR \simeq 
(1/2.14) d{\rm [Fe/H]}/dR$, so that $d$(O/H)$/dR \sim  -0.05\,$dex/kpc. Applying this value 
to the time variation of the gradients shown in Fig.\,\ref{fig3}, we have a maximum age of 
about 4 Gyr for the corresponding objects. From Table 1 of \cite[Bragaglia \& Tosi 
(2006)]{bragaglia}, all but one of the 17 open clusters listed have ages lower than 4 Gyr, in 
excellent agreement with our estimate. 

Some very interesting independent confirmation of the time flattening of the abundance 
gradients in spiral galaxies, and also of the flattening of the gradients at larger 
galactocentric distance arise from the recent work on M33 by \cite[Magrini et al. 
(2007a)]{magrini1} and \cite[Magrini et al. (2007b)]{magrini2}. In this work, data on HII 
regions, B supergiant and PN have been presented, and abundances of O, Ne, S and N have been 
studied. The derived O/H gradients are similar to the values obtained for the Milky Way, 
averaging about $-$0.10 dex/kpc for PN and $-$0.07 dex/kpc for HII regions, showing some 
time flattening during the last several Gyr in the life of the galaxy. The gradient in M33 
does not have a constant slope, and objects located further away along the disk show some  
evidences of a flattening gradient, similar to the one discussed here. In the scenario 
devised by the authors, the metallicity gradients flatten out with time as a result of the 
fact that chemical abundances at larger radii increase gradually with time, while a faster 
enrichment occurs  near the central regions. This result is particularly important, as it 
is the first evidence of a flattening in the abundance gradients in an external galaxy. 

\bigskip
{\bf Acknowledgements}. This work was partially supported by FAPESP and CNPq.

\end{document}